\documentclass{./aastex}
\usepackage{spr-astr-addons}
\usepackage{hyperref}
    \hypersetup{
    breaklinks=true
    }
\usepackage[]{natbib}

\RequirePackage{color}

\usepackage{amssymb}
\usepackage{graphicx}
\usepackage{epstopdf}

\begin{document}
\bibliographystyle{plainnat}
\title{History of Globulettes in the Milky Way }


\shorttitle{History of Globulettes in the Milky Way}
\shortauthors{Grenman, T., Elfgren, E., Weber, H.}

\author{Tiia Grenman\altaffilmark{1}} \and \author{ Erik Elfgren\altaffilmark{1}}
\and
\author{  Hans Weber\altaffilmark{1}}
\affil{Lule\aa\ University of Technology, Lule\aa, Sweden, SE-971 87 Lule\aa}
\email{Tiia.Grenman@ltu.se}

\begin{abstract}
Globulettes are small (radii $< 10$ kAU) dark dust clouds, seen against the background of bright nebulae. 
A majority of the objects have planetary mass.
These objects may be a source of brown dwarfs and free floating planetary mass objects in the galaxy. 
In this paper we investigate how many globulettes could have formed in the Milky Way and how they could contribute to the total population of free floating planets.
In order to do that we examine H-alpha images of 27 H~II regions. In these images, we find 778 globulettes.

We find that a conservative value of the number of globulettes formed is $5.7\times 10^{10}$.
If 10 \% of the globulettes form free floating planets then they have contributed with $5.7\times 10^{9}$ free floating planets in the Milky Way. A less conservative number of globulettes would mean that the globulettes could contribute $2.0\times 10^{10}$ free floating planets. 
Thus the globulettes could represent a non-negligible source of free floating planets in the Milky Way.

\end{abstract}


\section{Introduction}
\label{s:intro}

H II regions, surrounding young stellar clusters are driven by massive O and early B stars. These regions are associated with dusty gas formations, such as pillars, 'elephant-trunks'  and isolated dark clouds seen in different sizes and shapes, from large irregular blocks and fragments to smaller more roundish objects. These small-sized globules within H~II regions were first observed by \citet{bok47}, followed by  \citet{tha50} and \citet{her74}. They are dense, cold and neutral clouds with dark appearance in optical images. The pillars of gas and dust are pointing towards the ionizing sources and are usually connected to the molecular shell. 

In such H~II regions, a new distinct class of objects, globulettes, was noted by \citet{gah07} and \citet{gre06}. The globulettes are seen as silhouettes against the background of bright nebulae in optical images. Typically, they are roundish objects that are much smaller ($< 10$ kAU) than normal globules with a mass of  $\lesssim 0.1$~$M_\odot$ \citep{gah07}. The globulettes can also have tails, bright rims and halos. Many globulettes are quite isolated, located far from the molecular shells and dust pillars associated with the regions. However, some of the objects are connected by thin filaments to large molecular blocks or even each other. This suggests that they can either form separately, or via fragmentation of large structures. The calculated lifetime of globulettes suggests that they may survive in this harsh environment for a long time, $>10^6$~yrs  \citep{gah07,haw15}.

In the Carina Nebula (NGC 3372), a giant H~II region, the globulettes are smaller, more dense and less massive \citep{gre14} than those recognized in \citet{gah07} and \citet{dem06}. This group of globulettes may represent a more advanced evolutionary state in globulette evolution. \citet{schn16} suggested that the pillars may evolve first into a globule and then into a globulette when the lower density gas has photevaporated and left behind a dense and cold core. However, it is also speculated that globulettes might collapse in situ to form free floating planets (FFPs with $<13~M_{J}$, where $M_{J}=$ Jupiter mass) or brown dwarfs (13-75~$M_{J}$), triggered by external forces from gas and turbulent pressure in the surrounding warm plasma and radiation pressure from stellar photons.
In fact, the result by \citet{gah13} from near-infrared imaging of the Rosette Nebula, revealed very dense cores in some of the largest globulettes that might collapse to form FFPs. They also pointed out that these objects accelerate outwards from the nebula and eventually ends up in the interstellar environment. 

A more recent example, where globulettes may have formed planetary mass objects was found in the Orion Nebula \citep[M42/43,][]{fan16}. This nebula stands out by having both dark silhouette disks and bright compact knots. They are known as proplyds \citep{ode93, ode96, bal00}.

\begin{table*} [tbh]
\small
\centering
\caption{Archival HST / HLA / NOT data used}
\label{om}
\begin{tabular}{@{}lll@{}}

\tableline
H II region&Instrument&ID \\
 \tableline
Gum 29&ACS/WFC& 13038\\
Gum 31&WFPC2, ACS/WFC&10475, 9857\\
Gum 38b&WFC3/UVIS&11360\\
IC 434&WFPC2, ACS/WFC&8874,  9741, 12812\\
IC 2177&ALFOSC&N/A, see \cite{gah07}\\
IC 2944&WFPC2&7381\\
M 8&WFPC2, ACS/WFC &6227, 11981, 9857\\
M 16&WFPC2, ACS/WFC &10393,13926, 9091, 5773\\
M 17&WFPC2, ACS/WFC&6574, 8992 \\
M 20&WFPC2&9104, 11121\\
M 42/43&WFPC2, ACS/WFC&5469, 9825,10246  \\
NGC 281&WFPC2, ACS/WFC&8713, 10713, 9857 \\
NGC 1977 &ACS/WFC&12250\\
NGC 2174 &WFPC2, ACS/WFC&9091,13623, 9091 \\
NGC 2467&WFPC2, ACS/WFC&9857\\
NGC 3372&WFPC2, ACS/WFC& 6042, 11501, 13390, 13391, 13791, 10241, 10475\\
NGC 6357&WFPC2, ACS/WFC&9091, 9857\\
NGC 6820&ALFOSC&N/A, see \cite{gah07}\\
NGC 7635&WFPC2, ACS/WFC, WFC3/UVIS&7515,14471\\
NGC 7822&ALFOSC&N/A, see \cite{gah07}\\
S106&WFPC2, WFC3/UVIS&5963, 12326\\
S155&WFPC2&5983\\
S190&ALFOSC&N/A, see \cite{gah07}\\
S199&ALFOSC&N/A, see \cite{gah07}\\
S273&WFPC2, ACS/WFC, ALFOSC&8992, 5983\\
S277& WFPC2, ACS/WFC&5983, 8992, 9424\\
Rosette Nebula&ALFOSC&N/A, see \cite{gah07}\\
30 Doradus&ACS/WFC, WFC3/UVIS&11360, 12939\\
\tableline
\end{tabular}

\end{table*} 
 \begin{figure*}
\epsscale{2}
\plotone{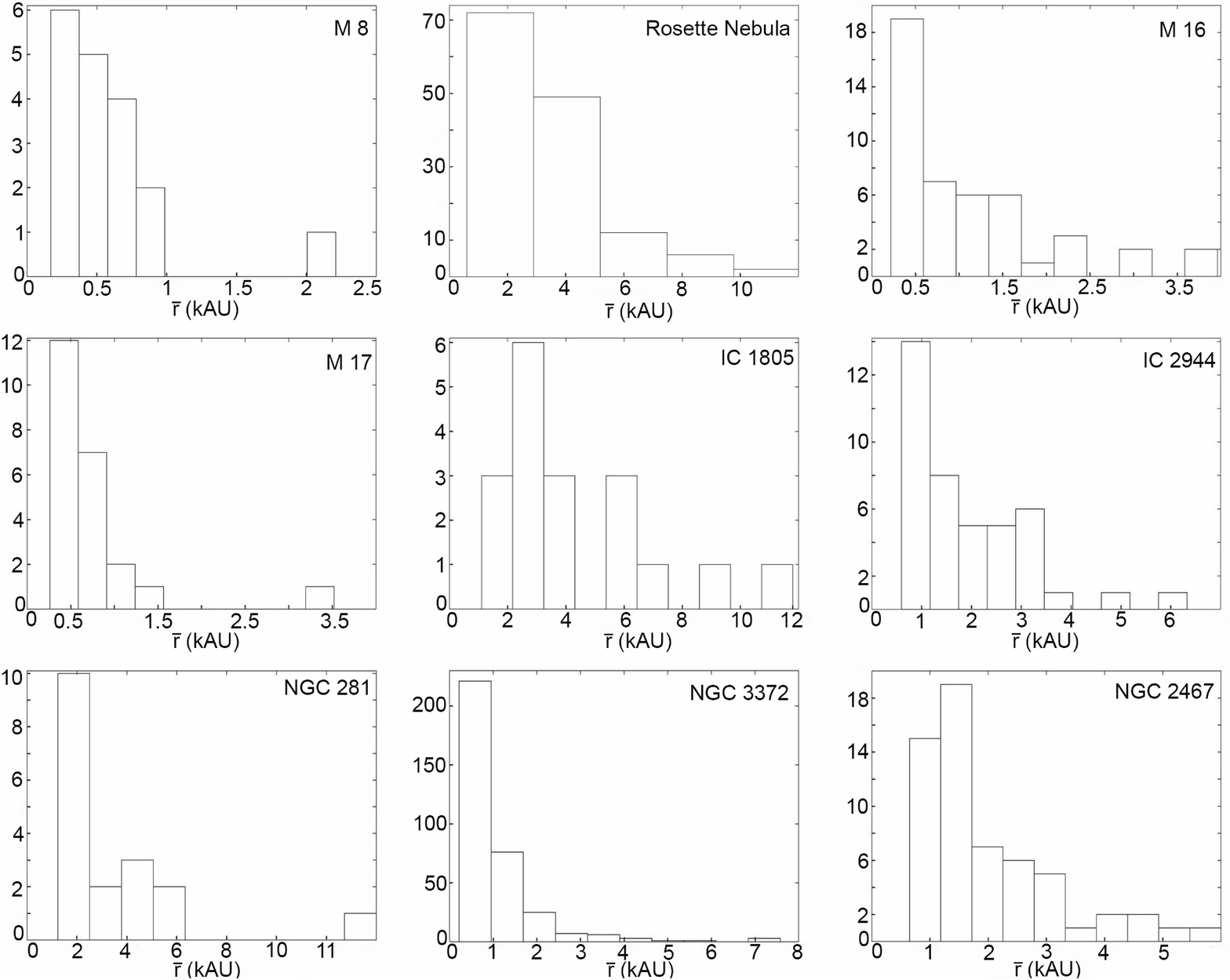}
\caption{ The mean radii distribution for those H II regions containing most globulettes. The histograms peak  at about 0.3~kAU in M 8, 2.5 kAU in the Rosette Nebula, 0.4 kAU in M~16 and M 17, 2.5 kAU in IC 1805, 0.9 kAU in IC 2944, 1.9~kAU in NGC 281, 0.6 kAU in NGC 3372 and 1.5  kAU in NGC 2467.}
\label{figA}
\end{figure*}
The first reports of populations of substellar objects in interstellar space came in 1995 \citep{reb95,nak95}. Since then, a number of Jupiter-mass FFPs and FFP candidates have been found in young clusters and star formation regions \citep[see][and references therein]{luc00, bih09, luh12, liu13, pen16, pen12, cla17}. Planetary mass objects have also been found in the galactic field \citep{cus14}. Since these substellar objects are both brighter and warmer at young ages \citep[less than 10~Myr,][]{cha00}, these objects with masses of a few Jupiter masses can be seen e.g. in deep optical and near-infrared photometric surveys \citep{zap00, zap17, qua10}. 

Another method to find objects below the deuterium-burning-mass-limit (Jupiter-sized) that are not bound to a host star is through gravitational microlensing. This type of study has been done by the Microlensing Observation in Astrophysics group \citep[MOA-2;][]{sum03, sum11,fre15} and the Optical Gravitational Lensing Experiment \citep[OGLE-III;][]{ wyr15}.

The origin of these unbound objects is unclear. They may have formed in situ from a direct collapse of a molecular cloud by fragmentation, similar to star formation \citep[e.g.][]{sil77, bow11}.  
In fact, \citet{fan16, luh05, bou16, joe15, bay17} have found four FFPs with accretion disk.
Furthermore, planetary-mass binaries, which would be almost impossible to from through planetary formation mechanisms,  have  also be found by e.g. \citet{jay06, bes17}.

FFPs may also have been formed from stellar embryos that are fragmented and photoevaporated from nearby stars \citep[see e.g.][]{kro03, pad04, whi04}. They may also have been ejected after they were formed in circumstellar protoplanteray disks through gravitational instabilities  \citep[e.g.][]{bos09, sta09, li15}. Such instabilities can be caused by a triplet star system \citep{rei15} or by passage through the center of a dense star cluster \citep {wan15}.

In this work, we present results based on available images of  H~II-regions from the Hubble Space Telescope (HST) archive and taken with the Nordic Optic Telescope (NOT). We create a model of how many globulettes may have formed in the Milky Way over time and how many of these that contribute to the total FFPs population. We have organized the paper as follows. In Sect. \ref{s:arhiv}, we present the result of both old and new H II regions investigated in the context of globulettes. We also present mean radii of histograms of globulettes for those H II regions, which contain many globulettes. In Sect. \ref{s:model}, the parameters that are used in our model are presented. The results and discussion regarding FFPs can be found in Sect. \ref{s:res} and we conclude the paper in Sect. \ref{s:con}.

\setlength{\tabcolsep}{4.3pt}
\begin{table*} [tbh!]
\small
\caption{ List of H II regions investigated} 
\begin{tabular}{@{}lrrlccccccc@{}}
\tableline
     
H II region &RA&Dec& Main Cluster&Distance & Ref.& Nebula & Obs.&Number&Detection \\
&(J2000.0) &(J2000.0)&&(kpc)&&Area&Area&of&limit\tablenotemark{**}\\
&&&&&&arcmin$^{2}$&arcmin$^{2}$&globulettes&(AU)\\

\tableline
Gum 29&10 24 & -57 46& Westerlund 2& 4.2 &1&707&11& 10&630\\
Gum 31& 10 37 & -58 39 & NGC 3324 & 2.3 & 2& 177&43& 7&345\\      
Gum 38b&11 15 & -61 15 & NGC 3603& 6.0 & 3& 94& 7& 5&720\\

IC 434&	05 41 &-02 30 &Sigma Orionis & 0.4 &4 &471 &86 &- &60\\
IC 2177\tablenotemark{*}&07 04 &-10 27 & N/A &1.3  &5& 314&33 & -&733\\
IC 2944&11 38  &-63 22&Collinder 249& 2.3&  6& 1\,866& 5& 34&690\\

M 8&18 03  &-24 23 & NCC 6530, 6523& 1.3 & 7&1\,374& 43& 18&195\\    
M 16&18 18  &-13 48 & NGC 6611&  1.8 &7&825& 53& 46&270\\
  M 17&18 20& -16 10  & NGC 6618&  2.0 &7& 806& 16& 23&300\\
 M 20&18 02  &-22 58 &NGC 6514& 2.7 & 7& 361& 21&-&810\\ 
M 42/43& 	05 35 & -05 23 & NGC 1976, 1982& 0.41 &7&1\,689&639&9& 61\\

NGC 281&	00 52 & +56 34 & IC 1590 &2.8&8 & 589 &32& 18 &420\\
 NGC 1977&05 35 	&-04 49   & NGC 1973, 75&0.41&9 &314& 44 &-& 61\\
NGC 2174&	06 09  &+20 30 &NGC 2175, 75s & 2.2  &12&589 &35 &15 &330\\  
 NGC 2467&07 52  &-26 25 & Haffner18ab,19 &5.0 &10& 467&16&59 &750\\

NGC 3372&	10 44 & -59 53 & Tr14/15/16,& 2.9  &11& 10\,179&953&343& 435\\
&&&Bo10/11,Cr232,\\
&&&Cr234/228\\

 NGC 6357&	17 26 & -34 12  &Pis 24, ESO 392-11 &1.7  &7& 911 &17 &11 &255\\
 &&&ESO 393-13 \\

NGC 6820\tablenotemark{*}&19 42 & +23 05 &NGC 6823&1.9 &13&707&53&1&1072\\
    
NGC 7635&	23 20  &+61 12 &[BDS2003]44 & 2.4 &14& 94& 37&4& 288\\

NGC 7822\tablenotemark{*}&00 01 & +67 25  &Berkeley 59&0.91&12  &1\,884& 72&9&513\\
S106& 20 27 	 &+37 22 & S106 IR&1.7 &15 &3&13  &-& 204\\
S155& 	22 58 &+62 31& CepIII& 0.84 &12 & 1\,178& 10 &-& 252\\

S190\tablenotemark{*}&02 32  &+61 27 & IC 1805 & 2.0 &12 & 5\,281& 189&18&1128 \\

S199\tablenotemark{*}&02 51 & +60 24& IC 1848, Cl34& 1.8 &12 & 5\,089& 66& -&1015\\
S273\tablenotemark{*}& 	06 41 &+09 53  &NGC 2264&0.91&7 &2\,672& 126 &- &136\\  
S277&	05 42 &-01 55& NGC 2024 &0.41& 7&491&107&3&61\\

Rosette Nebula\tablenotemark{*}& 06 31 &+04 57  &NGC 2244& 1.4 &16&5\,281&570&145&790\\

\tableline

LMC:&&&&&&& \\
30 Doradus& 06 31  &+04 57 & NGC 2070& 50 &17 &176&132&176&2\,000\\ 
\tableline  
    \end{tabular}
\tablenotetext{*}{H~II regions from \citet{gah07}}   
 \tablenotetext{**}{The minimum detection size (diameter in AU) of an object, defined as 3 pixels in this article.} 
\tablerefs{(1) \citealt{2013AJ....145..125V}; (2) \citealt{ohl13}; (3) \citealt{bra00}; (4) \citealt{moo09}; (5) \citealt{bic03}; (6) \citealt{san11}; (7) \citealt[and reference therein]{fei13}; (8) \citealt{sat08}; (9) \citealt{men07};
(10) \citealt{yad15}; (11) \citealt{hur12}; (12) \citealt {fos15}; (13) \citealt{kha05}; (14) \citealt{moo02}; (15) \citealt{sch07}; (16) \citealt{hen00}; (17) \citealt{pie13}  }

\label{regions}
      \end{table*}
\section{Archival Data and Measurements}
\label{s:arhiv}

In this work, we used archival H-alpha images of 16 H~II regions included in earlier studies by \citet{dem06}, \citet{gah07} and \citet{gre14}. The observations in these studies are from the HST archive and the NOT. Most images were taken with the HST instruments such as ACS/WFC, WFPC2, WFPC3/UVIS, while the NOT images were taken with the ALFOSC instrument. A brief description of these instruments is given below. 
\begin{enumerate}
\item The Wide Field and Planetary Camera 2 (WFPC2) field of view is covered by four cameras, each of which span $800 \times 800$ pixels in size. Three of them are arranged in an L-shaped field and operated at a pixel scale of $ 0.1''$. The fourth one is referred to as the Planetary Camera (PC) and operates at a pixel scale of  $ 0.046''$.  
\item  The Wide Field Camera 3 (WFC3) replaced the WPFC2 camera in 2009. The Ultraviolet-Visible channel (UVIS)  use a mosaics of two CCDs, with 0.04 arcsec/pixel, covering a $162'' \times 162''$ field of view. The 2 CCDs are butted together but have a $\sim 1.4''$ gap between the two chips.
\item The Advanced Camera for Surveys (ACS/WFC) camera contains two CCDs of $2048 \times 4096$ pixels glued together with a small gap of approximately 50 pixels in between. The pixel size corresponds to $0.05''$ per pixel and the field of view is $202'' \times 202''$.

\item The Andalucia Faint Object Spectrograph and Camera (ALFOSC) has a field of view of 6.5', and the effective scale of the CCD detector is 0.188'' per pixel.
\end{enumerate}

The respective areas for the instruments WFPC2, WFC3, ACS and ALFOSC are estimated to 5, 7, 11 and 33 arcmin$^{2}$. More information about each of these instruments can be found in the manuals on the HST Web site\footnote{\url{http://www.stsci.edu/hst/HST_overview/instruments}} and in the NOT manual\footnote{\url{http://www.not.iac.es/instruments/alfosc}}. All images in the current study are analyzed using the F656N and F658N filters, which allow for mapping circumstellar matter and detecting the dark silhouette objects, even though the F658N filter includes more nebular emission from [N~II].
The HST and NOT observations from these regions are presented in Table \ref{om}.

We characterize the angular size of the globulettes by using the SAOImage DS9 program, where we fit an ellipse to each object. Then we estimate the radius $\bar r $, the mean of the semi-major and semi-minor axes, expressed in kAU. 
In order to get a representative size distribution, a certain number of globulettes are needed. We have set the cutoff at 17 globulettes per region, and made 
histograms. Figure \ref{figA} shows the distributions of average radii in kAU for these H~II regions. The average radii distribution of the all globulettes seems to fall off approximately exponentially.

Using catalogues, such as \citet[hereafter Sharpless = S]{sha59} and \citet{gum55},
we have investigated 319 optically visible nebulae and found 28 H~II regions that have been imaged using an H-alpha filter. 
These regions are listed in Table \ref{regions}, Column~1, in alphabetical order with the coordinates (Columns 2 and 3) from the Simbad database\footnote {The SIMBAD database, operated at CDS, Strasbourg, France }. The main cluster designation is given in Column~4, and the distances from the Sun is given in Column~5, and the references in Column~6.  
The calculation of the H~II nebula areas in Column~7 is described in Sect. \ref{ss:neb}
and the calculation of the observed areas in Column~8 is described in Sect. \ref{ss:obs}. The next to last Column in Table ~\ref{regions}, lists the number of globulettes found in the observed area. In the last Column the minimum detection size of an object (in AU) is given. Objects with dimensions ${<}$ 3 pixels across are too small to be of interest \citep{dem06}.
For the NOT observations, there are no objects with a diameter of less than 9 pixels, which means that they are also clearly resolved.
 The Carina Nebula, Rosette Nebula and 30 Doradus contain the most globulettes. 30 Doradus is an example of a low metallicity region in the Large Magellinic Cloud (LMC) and we do not consider it in our calculation of globulettes within the Milky Way.

Some of the nebulae are giant H~II regions, like Carina Nebula  and 30 Doradus. In these regions, the Lyman continuum flux is $>10^{50}$ photons per second \citep{moi11}. There is also a young ultracompact H~II region, S106 \citep{cro03} in our list, where the ionized star is a pre-main sequence star embedded in its molecular cloud \citep{noe05}.
Most of the regions are relatively nearby ($d < 4$~kpc) and they are relatively young ($\lesssim$ 5~Myr).
The distances are based on the references cited below in Table \ref{regions}.

\section{Model Parameters}
\label{s:model}
In this section we define the different parameters needed to model the number of globulettes as shown in Table~\ref{parameter}. To illustrate the methodology of this work we chose the Rosette Nebula as a representative example, because it contains only one main cluster, is symmetric and hold a large amount globulettes.  An overview of the Rosette Complex can be found in \citet{rom08}.

\subsection{Young Universe and H~II Formation Rate} 
\label{ss:metal}

From model calculations we know that within the first relic H~II regions, the massive metal free Population III (Pop III) stars evacuated the primordial gas from their minihalos \citep{kit04, alv06, abe07}. As the Pop III stars ended in supernova explosions, the Universe's intergalactic medium got progressively enriched with heavy metals \citep[i.g.][]{gre07, wis08} and dust.

Cosmological simulations suggest that once the gas reaches the minimum or `critical' metallicity in the range $10^{-6}$-$10^{-3.5}Z_{\odot}$ which allows for more efficient cooling. An example of a low metallicity Pop~II environment is the Large Magellanic Cloud  (LMC), where the metallicity is  $\sim$ 0.5 $Z_\odot$ \citep{rol02}.

In \citet[Figure 11]{som15} these different metallicity values correspond to $T_0\sim 12$ Gyr, which we set as our starting point for the calculations of the number of globulettes in the Milky Way.

In the early Universe, the formation rate of the H~II regions was higher than it is now. We assume that it is proportional to the Star Formation Rate (SFR), which combined with the present day gives us the formation rate of H~II regions in the early Universe. \citet[Figure 1]{mar15}, show an overview of the SFR evolution as a function of time, where the current SFR is about $\sim 5$~$M_{\odot}$/yr. We estimate the total area under the curve, which gives us the average SFR between now and 10 Gyrs ago, which is twice as big as the H~II production is today. We call this correction factor $f_{H II}$, see Table~\ref{parameter}.

\subsection{Shapes and Areas of H II Regions}
\label{ss:neb}
H~II regions vary in their morphology and the nebulae examined here are either roughly circular, elliptical or irregular.
The angular sizes for each H~II region were measured from large scale DDS images (POSS II/UKSTU) taken through the red filter, with a plate scale of $\sim 1$'' per pixel, available in Aladin, an interactive software Sky Atlas\footnote{http://aladin.u-strasbg.fr/aladin.gml }. Thereafter, the area, in arcmin$^{2}$, for each H~II region was estimated, see Table \ref{regions} (Column 7), where the radii were defined as the geometric mean, $\sqrt{ab}$, where $a$ and $b$ are half the apparent major and minor axes. Here, the shape of the Rosette Nebula is assumed to be circular with an area of 5\,281 arcmin$^{2}$. The total area of the 27 H~II regions was estimated to $A_{tot}$ = 44\,413 arcmin$^{2}$.

 \subsection{Observed Areas of the H II Regions}
\label{ss:obs}
The DDS images of the 28 H~II regions from the HST archive AstroView were overlaid with the HST apertures while the survey done with the NOT is described in \citet{gah07}. The observed area (Column 8 in Table \ref{regions}) was calculated in arcmin$^{2}$ for each nebula. For the Rosette Nebula the area was estimated to be 570 arcmin$^{2}$ and the total observation area of the 27 regions, $A_{obs}$ is about 3\,297 arcmin$^{2}$.

\subsection{Objects behind the H II regions}
\label{ss:caviti} 

We can only detect globulettes in front of the H~II regions, while objects on the remote side are covered by nebular emission or hidden behind the foreground obscuring shell. Therefore, we assume that the backside of the nebula contains as many globulettes as we observe in front of the nebula. This gives a correction factor of $f_{B}=2$.

 \subsection{Age Estimate of the H~II Regions and Globulettes}
\label{ss:age}

An H~II region lasts only as long as the main sequence lifetimes of its ionizing star(s), i.e. 2-4 million years, before the radiation pressure from the hot young stars eventually drives most of the gas away and the nebula disperses. However, H~II regions can live longer either in large star forming regions, involving multiple clusters, progressive star formation (e.g. the Carina Nebula,  \citealt{smi10}) or when there is a spread in stellar ages, with co-existing younger ($\sim$ 1 Myr) and older  ($\sim $10~Myr) stars.  Two examples are Gum 38b \citep{bec10} and M 16, which have cluster mean ages of about 2 Myrs but where one of the massive members is about 6 Myrs old \citep{hil93}. Thus by taking these conditions into account, the mean life time, $T_{H II}$, for H~II regions is assumed to be 5~Myrs. 

The lifetimes of globulettes can be estimated to 4~Myrs, see Sect.\ref{s:intro}.

\subsection{Spatial Resolution and Distance Correction}
\label{ss:spa}

\begin{figure}
\epsscale{1}
\plotone{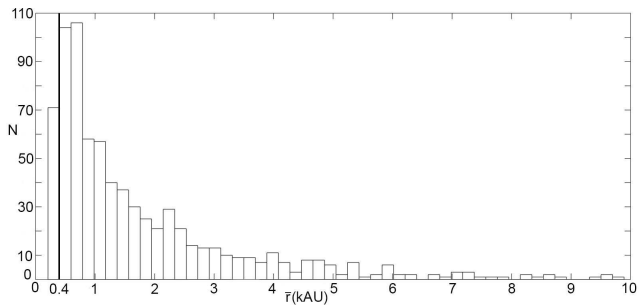}
\caption{ Distribution of average measured radii for all globulettes found in the H II regions, expressed in kAU. The detection limit of 0.4 kAU is also marked in figure.}
\label{figB}
\end{figure}

The spatial resolution of the ground based instrument NOT is lower than that of the HST. The practical globulette observation limit for the NOT telescope is about 0.8'', whereas the observation limit for the HST is about 0.15''. 
Thus we need to compensate for the objects that are below the resolution treshold for the NOT telescope.
A histogram of the distributions of radii observed with the different instruments is shown in Figure \ref{figB}. The most significant NOT region is the Rosette nebula, which has an observation limit of $r=0.8''\times 1.4\textrm{ kpc}=1.12$~kAU. The are about as many objects above this limit in the figure as there are objects below.
Therefore, the correction factor for the Rosette nebula is $f_{S}\approx 2$. For the other NOT regions, the correction factor would be only slightly different.
An approximate value for the total number of globulettes within the NOT regions is $G_{NOT}$=$f_{S}\times 173= 346$.
Furthermore, some H~II regions lie closer to us than others which means that we need to compensate for the missing objects at large distances. From Figure \ref{figB}, we see that the detection limit radii is $\sim 200$~AU for the more nearby regions. The only regions that are further away are NGC 2467 and Gum 38b. If we assume that doubling the distance, doubles the detection limit, the limit would be at 400~AU. In Figure \ref{figB}, we see that this correction between 200 to 400~AU
would correspond to 12\% if we assume that the distribution is similar. This factor also only applies to the two regions further away. Thus, this correction factor is negligible for the current data set.
 
\begin{figure}
\epsscale{1}
\plotone{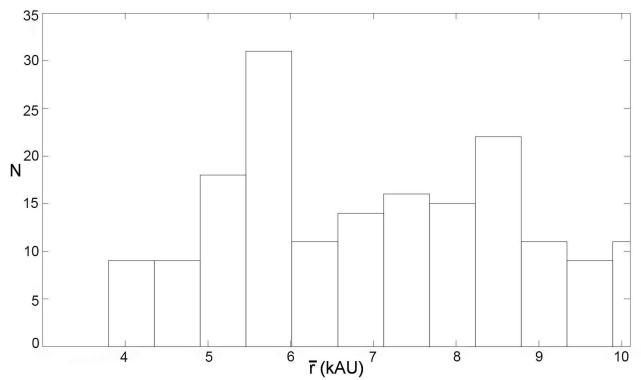}
\caption{ Size distribution of the 176 globulettes in 30 Doradus in terms of the mean radii. The distribution peaks at $\sim 5.7$~kAU. A few larger objects ($>$ 10 kAU) where not included in this study.}
\label{figC}
\end{figure}

\setlength{\tabcolsep}{6pt}
 \begin{table*} [tbh!]
\centering
\caption{Numerical values used in the calculations}
\label{parameter} 
\begin{tabular} {@{}lrll@{}}
\tableline
  Parameters/ &Numerical &Description& Section\\
 Symbols &Values&\\
\tableline  
$N_{H II}$&27& Number of analyzed Galactic H II regions& Table \ref{regions}\\
$T_{0}$&12 &Starting point in time for the model (Gyrs) &\ref{ss:metal}\\
$f_{HII}$&2& Correction for more H II regions in the early Milky Way&\ref{ss:metal}\\
$A_{tot}$&44\,413& Total area of 27 H II regions (arcmin$^{2}$) &\ref{ss:neb}\\
$A_{obs}$&3\,297& Total observed area of 27 H II regions (arcmin$^{2}$) &\ref{ss:obs}\\
$f_{B}$&2&Correction factor for globuletter on the backside of the nebula&\ref{ss:caviti}  \\
$T_{H II}$&5 &Lifetime of H II regions (Myrs)&\ref{ss:age} \\
$T_{glob}$&4 & Lifetime of a globulette (Myrs)&\ref{ss:age} \\
$f_{S}$&2&Correction factor between HST and NOT instruments&\ref{ss:spa}\\
$G_{NOT}$&346& Corrected number of globulettes observed with NOT&\ref{ss:spa} \\
$G_{HST}$&605& Total number of HST globulettes&Table \ref{regions}\\
$n_{H II}$&10\,000&Number of H II regions (present time)&\ref{ss:surveys}\\
$G_{Tot}$&951& Total corrected number of globulettes&\ref{s:res}\\
\tableline
\end{tabular}

\end{table*}  
\subsection{Number of H II Regions in the Milky Way} 
\label{ss:surveys}

H~II regions evolve with time; from hypercompact \citep[linear size $< 0.01$~pc; e.g.][]{kur02} to ultracompact \citep[$<0.1$~pc; e.g.][]{kim03} to compact \citep[0.1-1~pc][]{woo89} and then finally to extended H II~regions \citep[$> 1$~pc;][]{mel06}.
The evolutionary stage of a region is inferred from observations in several spectral windows. The most complete catalog today is compiled by \citet{and14, and15}, where about 8\,000 H~II regions and H~II region candidates in the Milky Way are registered. A round figure on the total number of H~II regions can be 10\,000 (L. D. Andersson, personal communication, 2016).
 
 \section{Results and Discussion}
\label{s:res}

In this section, we establish a model equation for the number of globulettes that have been formed in the history of the Milky Way. We also discuss how many FFPs may form from globulettes by taking account different parameters. In addition, we also discuss briefly the role played by giant H~II regions, like the Carina nebula and 30 Doradus, for producing globulettes in the early Milky Way.

In our scenario, the first globulettes were formed roughly 12~Gyr ago, when there was enough metallicity to create dust, (cf. Sect. \ref{ss:metal}). We estimate that H~II regions were 2 times more abundant in the early Milky Way than they are now (cf. Sect. \ref{ss:metal}) and that the number of H~II regions,  $n_{H II}$, today is about 10\,000 (cf. Sect. \ref{ss:surveys}). Globulette production can take place when the H~II region is still young, $\lesssim$5~Myrs
(cf. Sect. \ref{ss:age}) and we assume that we can find a similar amount of objects also on the backside of the nebulae (cf. Sect. \ref{ss:caviti}). The lifetime of globulettes was estimated to 4 Myrs.
The spatial resolution was lower in seven of the H~II regions and according Sect. \ref{ss:spa}, the actual number of globulettes, present in these regions, was estimated to 346 objects.
 
We list 27 H~II regions, where the total area was estimated to 44\,413 arcmin$^{2}$ and an observed area of 3\,297 arcmin$^{2}$. The total number of gloublettes was calculated as $G_{Tot}=G_{NOT}+G_{HST}=951$ and the average number of globulettes per H~II region was estimated to be about 35. According to our model, the number of globulettes that have formed in the history of the Milky Way (MW) is
\begin{equation} 
G_{MW}=\frac{T_{II}}{T_{glob} }\cdot \frac{T_{0}}{T_{H II} } \cdot  n_{H II} \cdot f_{B} \cdot  f_{H II}\cdot \frac{A_{tot}}{A_{obs}} \cdot \frac{G_{tot}}{N_{H II}}.
\label{eq:1}
\end{equation}
Using the numerical values from Table \ref{parameter} we get a total number of globulettes $5.7\times 10^{10}$ formed in the Milky Way over time, which can be considered a conservative value. 
It is currently not known what fraction of the  globulettes that form FFPs. Their mass and density are likely to affect this fraction but to what extent is currently unknown.
Neverthless, if we assume that 1 \% or 10 \% of the globulettes form FFPs, then the number of FFPs originating from globulettes is $5.7\times 10^{8}$ and  $5.7\times 10^{9}$, respectively.

The MOA-2 Galaxtic bulge microlensing survey \citep{sum11} found that the Jupiter-sized FFPs are about 1.8 times as common as main-sequence stars.
These planets are unbound or bound but very distant $>$100 AU from any star. It is unclear if some of these objects are low-mass brown dwarfs or super-Jupiters. The difference between these classes of objects is not well defined  \citep{luh12, gia13}. However the population of these objects is expected to arise from a variety of processes, such as in situ formation or from the core accretion theory, see \citet{ver12} and Sect. \ref{s:intro}. 
However, the number of Jupiter-sized FFPs by \citet{sum11} may be overestimated, due to blending or red noise in the data \citep{bac15} or the distribution of the timescale of mircolensing events \citep{dis12}.
Furthermore, \citet{cla17} also found a slightly lower value ($1.2$-$1.4$) of FFPs per main sequence star. Recently, \cite{nature} found, using microlensing, 0.25 per star as an upper limit on the number of Jupiter sized planets.

A more conservative estimate of the number of Jupiter-sized planets has been calculated by \citet{tut12} and more recently by \citet{ma16}, where the FFP population is found to be $\sim 1.8 \cdot 10^{-3}$ of the stellar population, which corresponds to $\sim 10^8$ a total number of FFPs. 
However, these FFPs candidates originate in protoplanetary disks, while the FFPs considered here are formed in situ.
Their value is less than the observed value of \citet{sum11} but it is comparable to our result if 1\% of the globulettes form FFPs.

If we increase the factors $f_{HII}=2\rightarrow 3$, $f_s=2\rightarrow 5$ and $f_B=2\rightarrow 3$, equation~\eqref{eq:1} would yield about $2.0\times 10^{11}$ globulettes. Then by assuming that 10 \% of the globulettes form FFPs, the globulettes would then contribute to about 11 \% of all FFPs, as estimated by \citet{sum11}.

A number of candidates of brown dwarfs and Jupiter mass objects have been found in several nearby, young star clusters and star-forming regions. \citet{dra16} have found 160 isolated planetary mass object candidates in the Orion Nebula. However, we only observe 9 dark objects in this H~II region. This can be compared to the Rosette Nebula, where we have 16 times more objects in a smaller area.

We note that the total observed area of the H II regions is only 7 \% of the total nebula area, and in the 27 H~II regions there are 18 nebulae with $<18$ observed globulettes and 8 nebulae with no observed globulettes.
In these nebulae, the surveys could simply have missed them or the enivronment might not be favorable for globulette formation.

In  30 Doradus,  which is the closest extragalactic source, many filaments and dust clumps in different sizes have been observed \citep{ind13, che16}. In this nebula, we discovered 176 globulettes with mean radii less than 10 kAU. Their size distribution is shown in Figure \ref{figC} with a peak at about $\sim$ 5.7~kAU. 
Compared to the globulettes in H~II regions within the Milky Way, these globulettes are rather large.
Due to its far distance, the detection limit of the objects in the 30 Doradus is about 4~kAU, which means that all smaller globulettes have escaped detection.

\section{Conclusions} 
\label{s:con}

In this paper we have examined 319 H~II regions, whereof 28 were observed with a narrowband  H-alpha filter. 
In these H~II regions, we have identified small roundish objects, globulettes, which appear dark in the H-alpha images.
We summarize the present work as follows;
 \begin{enumerate}
 \item The total area and observed area of the 27 H~II regions was estimated to 44\,413 armin$^2$ and to 3\,297 arcmin$^2$, respectively.
 
 \item We estimated the mean radii for each globulette found in the H~II regions. The majority of objects have radii $<10$ kAU but most seem concentrated around $\sim 0.5$~kAU.

 \item Amongst the 27 H~II regions, there are 18 nebuale with $<18$ observed globulettes and 8 nebula with no observed globulettes.

   \item  In the 30 Doradus, the mean radii had a peak at $\sim$ 5.7 kAU, which was higher than the regions within the Milky Way.

\item The total amount of globulettes was estimated to 778 and by correcting for the factor $f_s$ (resolution difference between HST and NOT), the number of globulettes was estimated to 951, which gives an average of 35 globulettes per H II region.

\item Our model gives a conservative estimate on the number of globulettes formed in the history of the Milky Way to about $5.7\times 10^{10}$. A less conservative estimate gives $2.0\times 10^{11}$ globulettes.

\item The globulettes could therefore represent a non-negligible source of the FFPs in the Milky Way. For example, if 10 \% of the globulettes of our less conservative estimate form FFPs, then the globulettes would contribute to about 11 \% of all FFPs, as estimated by \citet{sum11}.

\end{enumerate}

\acknowledgments
We wish to thank professor G\"osta Gahm for many fruitful discussions.
We also wish to aknowledge the data provided by the NASA/ESA Hubble Space Telescope (\facility{HST}), obtained at the Space Telescope  Science Institute
as well as the Nordic Optical Telescope (\facility{NOT}).

\end{document}